\documentclass[preprint,aps,amsmath,amssymb,superscriptaddress]{revtex4}  

\usepackage{graphicx}
\newcommand{\ev}[1]{\left\langle #1 \right\rangle}

\begin{document}

\title{On Modified Dispersion Relations and the Chandrasekhar Mass Limit}

\author{Michael Gregg} 
\affiliation{Department of Physics, Hamilton College, Clinton NY 13323
USA}
\author{Seth A. Major} 
\email{smajor@hamilton.edu}
\affiliation{Department of Physics, Hamilton College, Clinton NY 13323
USA}  

\date{July 2008}

\begin{abstract}   
Modified dispersion relations from effective field theory are shown to alter the Chandrasekhar mass limit.
At exceptionally high densities, the modifications affect the pressure of a degenerate electron gas and can increase or decrease the mass limit, depending on the sign of the modifications. These changes to the mass limit are unlikely to be relevant for the astrophysics of white dwarf or neutron stars due to well-known dynamical instabilities that occur at lower densities.  Generalizations to frameworks other than effective field theory are discussed.
\end{abstract}

\maketitle 

\section{Introduction}

The principle of Lorentz Invariance is at the heart of the contemporary formulation of physical theory, in particular the Standard Model and 
general relativity.  Given the centrality of Lorentz Invariance (LI) it is wise to explore as many avenues as possible that test this principle.  One avenue investigated in recent years is the physics of modified dispersion relations (MDR).  A remarkable result of these studies is that astrophysical data significantly limit Planck scale - suppressed modifications (see e.g. \cite{jlm_rev,d_rev} for reviews). The successful limits on the modifications are due to the high degree of sensitivity of particle process thresholds to Lorentz Violation (LV). Given the delicate interplay between the modifications of the particles involved, it is helpful to look for systems in which the effects of the modification on a single particle type are isolated.  An apparently ideal system is the physics of a degenerate electron gas since the pressure of such a gas supports the gravitational attraction of white dwarfs.  Further, the Chandrasekhar mass limit, about $1.4 M_\odot$, of white dwarf stars is obtained in the ultra-relativistic limit, precisely where modifications of dispersion relations are expected to be large. 

Modified dispersion relations often take the form of an expansion in LV terms
\begin{equation}
E^2 =p^2 +m^2 + \kappa_3 \frac{p^3}{M_P} + \kappa_4 \frac{p^4}{M_P^2} + ...
\end{equation}
where the parameters $\kappa_i$ may differ for different particle species and $M_P$ is the Planck mass, which we take to be $M_P = \sqrt{ \hbar / ( 4 \pi G)} \approx 3.45 \times 10^{27}$ eV ($c=1$).  In such MDR models the usual energy-momentum conservation laws hold so there is a preferred frame, which we take to be the one where the cosmic microwave background is isotropic\footnote{At the conclusion of the paper we comment on these results in MDR frameworks in which LI may not be broken.}.  While these modified dispersion relations could be viewed simply as a phenomenological expansion to test LV, this form has been suggested in a variety of settings including string theory tensor vacuum expectation values, heuristic calculations of the semiclassical limit of loop quantum gravity,  spacetime foam, non-commutative geometry, analogs of emergent gravity, and some braneworld models \cite{jlm_rev,d_rev}.  

Given the energy scale of the modifications, it might seem that testing such modifications might  simply be impossible,   However even the early work \cite{limits,jlm} demonstrated that particle process thresholds are highly sensitive to these modifications.  Using several particle processes and observed energies, much of the parameter space is ruled out \cite{limits,jlm}.  For instance the recent work of Macione {\em et. al.} achieves a limit on the parameter space of electrons  of less than $10^{-5}$ \cite{new_crab}. 

In this paper we report on numerical solutions to the exact equations for the Chandrasekhar mass limit with modified dispersion relations. We extend the analysis of \cite{camacho}, relaxing the unphysical assumption of constant density.  We find significant differences with the reported results: The mass limit may be raised or lowered depending on the sign of the modifications in the electron dispersion relation; and physical equilibrium radii exist for both signs of the MDR parameter $\kappa$. Finally we show that, despite the existence of new mass limits, there would be no effect on white dwarf astrophysics.  The effects are only important in the Planck-scale regime far above astrophysically accessible densities.

This paper is organized as follows.  The next two sub-sections are devoted to discussions of MDR and the mass limit.  In section \ref{mod_mass} we derive the corrections to the mass limit due to MDR.  In the final section \ref{conclude} we summarize the results, compare with results previously reported in \cite{camacho}, and comment on the applicability of the calculation to other frameworks with modified dispersion relations.

\subsection{Modified Dispersion Relations}

To achieve precise limits on the parameters it is necessary to have some additional knowledge of the dynamics of the field theories. This may be achieved in the context of effective field theory, where effective field theory is used to determine the (non-renormalizable) mass dimension five (or higher) LV operators.

Myers and Pospelov found that there are essentially only three operators that simultaneously break local LI and preserve gauge and rotation invariance.  Introducing a preferred frame field $n^a$, the mass dimension 5 operators are \cite{MP}
\begin{equation}
- \frac{\zeta}{M_P} n^a F_{ac}\, n \cdot \partial ( n_b \tilde{F}^{bc} ) + \frac{1}{2 M_P} n^a \bar{\psi} \gamma_a ( \xi_1 + \xi_2 \gamma_5)(n \cdot \partial)^2 \psi
\end{equation}
where $ \zeta, \xi_i$ are dimensionless parameters, $\tilde{F}^{ab}$ is the dual of the usual $F_{ab} = \partial_a A_b - \partial_b A_a$ of Maxwell theory.  The dispersion relations for fermions become
\begin{equation}
E_\pm = p^2 +m^2 + \eta_\pm \frac{p^3}{M_P}
\end{equation}
where the different signs correspond to the two helicity states.
Typically, the modifications become important when they become comparable to the mass terms.  For an electron this occurs at an energy scale of $\sqrt[3]{m_e^2 M_P} \sim$ 10 TeV - significantly below the Planck scale.  While these energy scales are beyond terrestrial experiments, astrophysical processes at these energy scales effectively constrain the cubic modifications \cite{jlm,new_crab,limits}.  

The fact that electron cubic MDR have two parameters leads to new effects. At sufficiently high energies electrons are unstable to helicity decay \cite{jlm_rev}.  If $\eta_-> \eta_+$ then the negative helicity electrons will decay producing a photon and positive helicity electron.  While there is no kinematic threshold for this process, an effective threshold may be derived by studying the reaction rate.  The resulting effective threshold is $(m_e^2 M_P/ \Delta \eta)^{1/3}$ or about 10 TeV for $\Delta \eta = |\eta_+ -\eta_-| < 0.4$, as determined by photon stability \cite{jlm_rev}.  The lifetimes of the negative helicity electron decrease around this effective energy; lifetimes for a 1 TeV electron is about 1 s while for a 50 TeV electron it is reduced to about $10^{-9}$ s \cite{jlm_rev}.  However the lifetimes increase at higher energies due to the fact that the states become more chiral. The MDR effects on the Chandrasekhar mass only occur at such high energies that the electron population is effectively in one helicity state. Assuming that the astrophysical processes, e.g. accretion, driving the increase in electron energy are sufficiently slow as compared to the lifetimes of the helicity states all the electrons would have positive helicity.  Since the same analysis applies for the other case,  $\eta_+> \eta_-$, the resulting degenerate electron gas will all be in the helicity state associated with the smallest parameter.  Hence we will denote the electron parameter with $\kappa$, where $ \kappa = \text{min}(\eta_+,\eta_-)$. 

For the purposes of studying the effect on the physics of compact stars we assume $\kappa$ is order one and write the MDR as
\begin{equation}
\label{mdr}
E^2 = p^2 \left[ 1 + \kappa \left( \frac{p}{M_P} \right)^n \right] +m^2.
\end{equation}
While the case of dimension 5 operators and cubic modifications ($n=1$) is the most interesting, since the analysis easily generalizes we present the mass limit calculation for arbitrary $n$.  The range of validity for this effective description is $p<M_P$.

\subsection{Chandrasekhar Mass Limit}

White dwarf stars are an end stage of stellar evolution for stellar masses less than $\sim 8 M_\odot$.   As the star cools and contracts, the electron wavefunctions begin to overlap appreciably. As the star contracts further, the electrons are forced into higher and higher momentum states to satisfy the Pauli exclusion principle, forming a  degenerate electron gas with relativistic Fermi energy. Because the thermal energy is so much lower than the Fermi energy the temperature is effectively zero.
The Chandrasekhar mass limit, about $1.4 M_\odot$,  of white dwarfs stars arises when the Newtonian gravitational attraction due to the nuclei (often comprised of carbon and helium) is balanced by the outwards pressure of degenerate, cold electrons. The mass limit is obtained for an ultra-relativistic Fermi gas, when $E/m_e \sim 10^4$ with central densities  in excess of $ \rho \sim 10^6$ g cm$^{-3}$.  
Thus the Chandrasekhar mass limit is derived in the framework of Newtonian gravitation and the statistical mechanics of an ideal, non-interacting, ultra-relativistic gas of electrons at $T=0$. 

The Chandrasekhar mass limit is largely of theoretical interest. Other physics comes into play at high densities introducing processes known as dynamical instabilities.  Depending on the details of the star's composition, two processes transform the physics of the star. At densities  $\rho \sim 10^9 - 10^{11}$ the electrons acquire sufficient energy to induce inverse $\beta$-decay and the star turns into a neutron star in a process called neutronization.  At the critical density of  $\sim 10^{10}$ g cm$^{-3}$ the onset of general relativistic instabilities causes the star to collapse.  These threshold densities are composition dependent so, for instance, the general relativistic instability is irrelevant to iron white dwarfs, which undergo neutronization before gravitational collapse.  For helium and carbon white dwarfs, the situation is reversed, instabilities due to general relativity occur before neutronization.

\section{Chandrasekhar Mass Limits with Modified Dispersion Relations}
\label{mod_mass}

To determine the actual mass limit - with or without modifications - a numerical analysis must be done.   
In calculating the electron degeneracy pressure, which supports the gravitational attraction, we assume that the star is isotropic and assume that the velocity of the star is not large in the preferred frame.  The pressure is given by
\begin{equation}
P = \frac{1}{3} \ev{n_e \, \boldsymbol{p \cdot v} }
\end{equation}
where $n_e$ is the electron number density.  We use the group velocity in the MDR case, so that 
\begin{equation}
v := \frac{d E}{d p} = \frac{p}{E} \left( 1 + \kappa \, \frac{n+2}{2} \frac{ p^{n}}{M_P^n} \right).
% same conventions as Mike
\end{equation}
As the temperature is effectively zero, all the states below the Fermi momentum, $p_F$,  are occupied.  Hence the pressure in the continuum limit simply becomes
\begin{equation}
\label{Pform}
P = \frac{8 \pi}{3 h^3} \int_0^{p_F}   \frac{p^4}{E} \left(1  + \kappa \, \frac{n+2}{2} \frac{ p^{n}}{M_P^n} \right) dp
\end{equation}
Introducing the dimensionless momentum $ x := p/m_e$ and the parameter $\delta := \kappa (m_e/M_P)^n$ we have
\begin{equation}
\label{Pxform}
P = \frac{8 \pi m_e^4}{3h^3} \int_0^{x_F} \frac{x^4 \left(1 + \delta \, \frac{\left(n+2\right)}{2}  x^{n} \right) }{\sqrt{1+x^2 + \delta \; x^{n+2} } } dx
\end{equation}

A white dwarf's mass depends on the star's composition. To express the white dwarf density then it is conventional to parameterize the density in terms of the number density of electrons and the atomic mass unit $m_u =1.66 \times 10^{-24}$ g, so that $\rho = \mu \, m_u \, n_e$, where $\mu$ is the electronic molecular weight.   Since white dwarfs are composed mostly of carbon with traces of helium, the proton-electron density ratio is close to $2$ since these nuclei have electron-nucleon ratios of 1 to 2.

Before setting up the numerical solution, it is useful to estimate the affect of the modifications.  From equation (\ref{Pxform}), one can show that  the pressure in the ultra- relativistic limit is approximately
\begin{equation}
P \simeq \left( \frac{2 \pi m_e^4}{3 h^3} \right) \left[ x_F^4 + \delta \frac{ 2 (n+1)}{(n+4)} x_F^{n+4} \right].
\end{equation} 
Thus, the MDR effectively add an attractive (for $\kappa<0$) or repulsive (for $\kappa>0$) force that scales with the Fermi momentum as  $p_F^{n+4}$. 

In equilibrium the gravitational attraction must be balanced by the outward force due to the pressure of the degenerate gas.  So, up to factors of order one, the ratio $GM^2/R^4P$ should be constant.  Using the pressure in the ultra-relativistic regime and $\rho \sim M/R^3$ we see that this ratio becomes 
\begin{equation}
\label{urlimit}
\frac{GM^2}{R^4 P} \propto M^{2/3} \left[ 1- \kappa \left( \frac{ M^{1/3} h}{m_u^{1/3} M_P R} \right)^n \right] ,
\end{equation}
neglecting factors of order one.  In the non-relativistic case the ratio scales as $M^{1/3}R$ so increasing mass requires decreasing radius to maintain the value of the ratio.  This drives the system into the relativistic regime given in equation (\ref{urlimit}). The leading term is constant in $R$ indicating that here are equilibrium solutions for suitable mass - the mass limit - and radius. However, if the mass exceeds the limit then the gravitational force will exceed the pressure and the star will collapse. The correction due to MDR modifies this description.  In equilibrium the ratio of equation (\ref{urlimit}) has a definite value, as before.  For $\kappa>0$ as the mass $M$ is increased, equilibrium may be restored by decreasing the star's radius $R$, achieving a new equilibrium at higher density.  For $\kappa<0$, as the mass is increased the correction term must play a role and the star must increase in relative size to remain in equilibrium. For a star at the Chandrasekhar mass limit the correction, approximately $(M/m_u)^{n/3} (L_P/R)^n$, is negligible, only $10^{-14}$ for $n=1$.  ($L_P$ is the Planck length.) We will see that this is born out in the numerical analysis.

At $T=0$ all states are inside a sphere of radius $p_F$ in momentum space so the number density is simply
\begin{equation}
n_e= 2 \frac{4\pi}{3}{\left(\frac{p_F}{h}\right)}^3
\end{equation}
giving a density of
\begin{equation} \label{dens}
\rho= \frac{8\pi}{3} \mu m_u {\left(\frac{m_e}{h}\right)}^3x_F^3 \equiv \mu \rho_o x_F^3
\end{equation}
where
\begin{equation}
\rho_o := \frac{8\pi}{3} m_u {\left(\frac{m_e}{h}\right)}^3 \simeq 9.81 \times 10^5 \text{ g cm}^{-3}
\end{equation}
% this is the B from Padmanabhan
and $x_F = p_F/m_e$.

In the context of the Chandrasekhar mass limit (see \cite{pad,ST}), the star is in equilibrium when the gradient of the degeneracy pressure supports  the Newtonian gravitational attraction.  For a spherically symmetric mass distribution the radial pressure gradient satisfies
\begin{equation}
\label{equilib}
\frac{ d P}{dr} = - \frac{G \rho(r) m(r)}{r^2}
\end{equation}
Differentiating and gathering terms gives
\begin{equation} 
\label{equilibrium}
\frac{1}{r^2}\frac{d}{dr}\left(\frac{r^2}{\rho}\frac{dP}{dr}\right)=-4 \pi G \rho.
\end{equation}
Note that we have not made the approximation that the density $\rho(r)$ is uniform.

To set up the numeric calculation it is useful to introduce the dimensionless electron energy $z(r) = E(r)/m_e$ and its value at the center of the star,  $z_c:=z(0)$.  A convenient dimensionless  radius $\zeta$ is given by \begin{equation}
\zeta := \frac{r}{r_o} \text{ with } r_o := \frac{1}{2 \sqrt{3}} \frac{M_P}{m_u} \frac{h}{ m_e} \frac{1}{\mu z_c}  
\simeq  7.77 \times 10^9 \frac{1}{\mu z_c} \text{ cm}.
\end{equation}
% this is the alpha from Padmanabhan
Normalizing the dimensionless energy $z(\zeta)$ to the value at the center $z_c$ gives the normalized energy $Q(\zeta) := z(\zeta) / z_c$.  The radial evolution of this normalized energy is determined by the equilibrium condition (\ref{equilibrium}).  To derive the resulting equation note that
\begin{eqnarray}
\frac{1}{\rho} \frac{dP}{dr} &=& \frac{1}{\rho} \frac{dP}{dx_F} \frac{dx_F}{dz}\frac{dz}{dr} \\ \nonumber
&=& \left( \frac{1}{\mu \rho_o  x_F^3}  \right) \left( \frac{8 \pi m_e^4}{3 h^3} \right) \left[  \frac{ x_F^4 \left( 1+ \delta \frac{(n+2)}{2} x_F^{n+1} \right)}{\sqrt{1+x_F^2 + \delta x_F^{n+2}}} \right] \left( \frac{z}{x_F + \delta \frac{n+2}{2} x_F^{n+1}} \right) \frac{dz}{dr}\\
&=&  \left( \frac{8 \pi m_e^4}{3 \rho_o \mu h^3} \right) \frac{dz}{dr}
\end{eqnarray}
The third line follows from the fact we used the group velocity in the expression for the pressure.  Using this result and the dimensionless radius $\zeta$ in equation (\ref{equilibrium}) gives  
\begin{equation}
\label{numready}
\frac{ d^2 Q}{d \zeta^2} + \frac{2}{\zeta} \frac{d Q}{d \zeta} + \left( \frac{x_F(Q)}{z_c} \right)^3 =0
\end{equation}
where $x_F(Q)$ is the dimensionless momentum in terms of $Q(\zeta)$ as determined by the equation
\begin{equation}
\label{xeqn}
Q^2 = \frac{1}{z_c^2} (1+ x_F^2 + \delta x_F^{n+2} ).
\end{equation}
In the usual context of special relativity when $\kappa = \delta =0$, $x_F(Q)$ is simply given by $ \sqrt{Q^2 - 1/z_c^2}$.  For large values of the central energy, it is possible to approximate the $x_F(Q)$ as
\begin{equation}
\label{approx}
x_F \simeq z_c \sqrt{ Q^2 - \frac{1}{z_c^2}} \left[ 1- \frac{1}{2} \delta z_c^n \left( Q^2 - \frac{1}{z_c^2} \right)^{n/2} \right]. 
\end{equation}
The boundary conditions for the evolution equation (\ref{numready}) at $\zeta=0$ are $Q(0) =1$, by definition, and $dQ/d\zeta |_{\zeta=0} = 0$, to ensure vanishing pressure gradient as required by equation (\ref{equilib}).

\begin{figure}
      \begin{center}
    \includegraphics[scale=1]{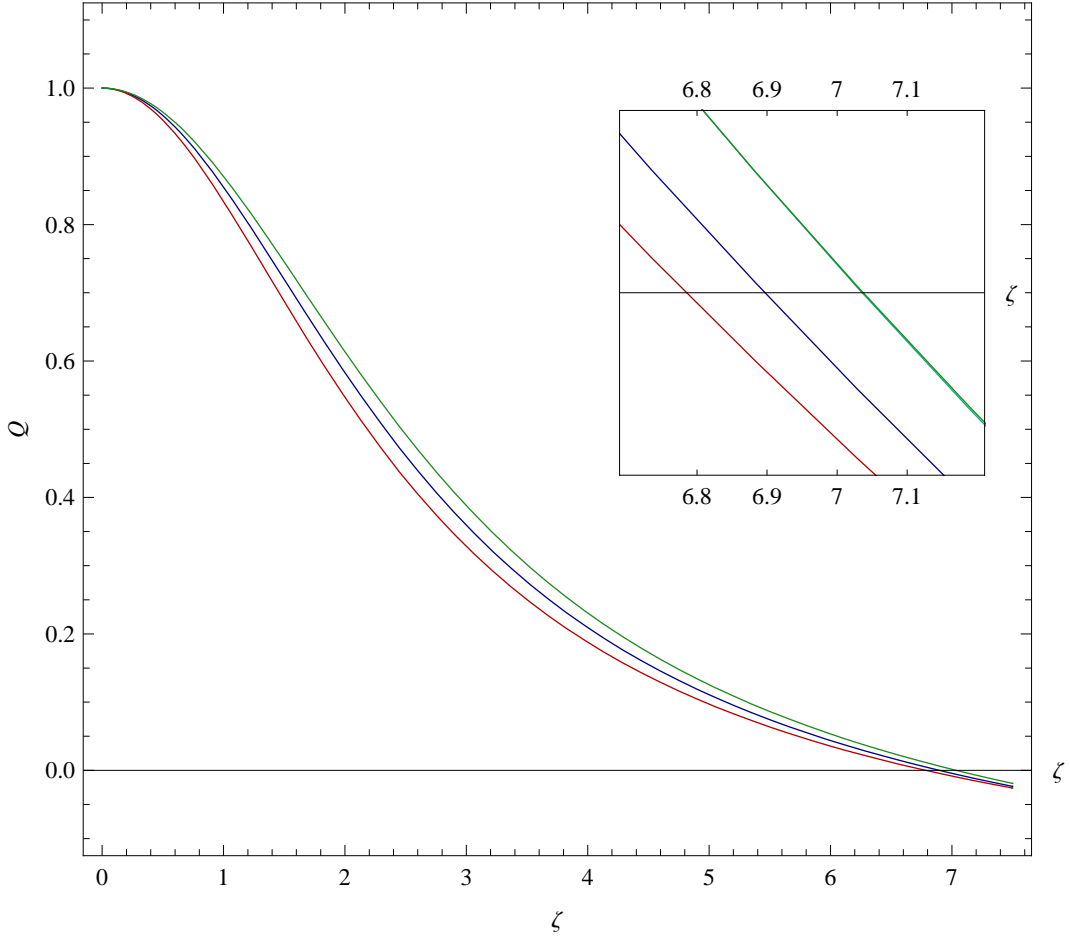}
  \end{center}
      \caption{\label{qresults} Results from numerical integration of the equilibrium 
      condition for $n=1$, $\kappa = \pm1$, and $z_c = 10^{21}$. Solutions for the normalized energy $Q(\zeta)$ are plotted for  $\kappa = -1$ the lower (red) curve, $\kappa=0$ - the unmodified case - the middle (blue) curve, and $\kappa = +1$ the upper (green) curve.  The inset plot shows the neighborhood of $\zeta =-7$.  This plot shows the zeros of $Q$ and the relative slopes, important for determining the limiting mass, for the cases $\kappa=-1, \kappa =0, \kappa = +1$, from left to right.  The color coding is the same as in the larger plot.  In addition, the numerical solution with approximation of equation (\ref{approx}) is included for the $\kappa=1$ case. At this scale it is barely distinguishable from the solution with the exact relation for the density.}
\end{figure}

The non-linear differential equation (\ref{equilibrium}) may be solved numerically, once values for the parameters $n, \delta$, and $z_c$ are chosen.  To find solutions, the normalized energy $Q$ is evolved via equation (\ref{numready}) outward from the center of the star.  The outer boundary is determined by vanishing density, which in terms of the normalized energy means that 
$x_F(Q) = 0$.  When $z_c$ is very large, this is well approximated by $Q(\zeta_R)\approx 0$.   For an example numerical solution consider the case $n=1$, $\kappa = -1$, and $z_c = 10^{21}$.  First the physical root of the cubic equation for $x_F(Q)$, equation (\ref{xeqn}), is found by selecting the root for which the density is positive and small for small $Q$ (since $\rho \propto x_F^2 \sim Q^3$).  Using this root the resulting equation (\ref{numready}) is numerically evolved outward until $Q(\zeta_R)= 0$. The resulting numerical solution is shown in figure \ref{qresults} as the (red) leftmost curve. The other solutions for $n=1$ are found in a similar manner.  The case for which $n=1, \kappa=+1$, and $z_c = 10^{21}$ is the (green) rightmost  curve.

 The mass of the star is determined by integrating the density to the surface of the star $r=R$ \cite{pad,ST}
\begin{eqnarray}
M &=& \int_0^R 4 \pi r^2 \rho(r) dr = 4 \pi \mu  \rho_o \alpha^3 \int_0^{\zeta_R} \xi^2 x_F^3(\zeta) d \zeta \nonumber \\
&=& 4 \pi \mu  z_c^3 \rho_o \alpha^3 \left[ - \int_0^{\zeta_R} \frac{d}{d \zeta} \left( \zeta^2 \frac{dQ}{d\zeta} \right) d\zeta \right] \nonumber \\
&=& \sqrt{3} \pi^2 \left( \frac{2}{\mu} \right)^2  \frac{M_P^3}{m_u^2} \left[ - \zeta^2 \frac{dQ}{d\zeta} \right]_{\zeta_R}
\end{eqnarray}
where the second line follows from equation (\ref{numready}).  Thus for large $z_c$, the product of the root of $Q(\zeta)=0$ and the slope at this radius give the mass of the star.  

The mass limit $M_{\rm ch}$ is determined in the ultra-relativistic limit.  This is often denoted $z_c \rightarrow \infty$.  However in the present case the onset of the MDR contributions to the pressure occur at central energies of $z_c \sim 10^{16}$.  Since this energy is so well separated from the usual scale $z_c \sim 10^4$, we can clearly see the regime of the usual Chandrasekhar mass limit and a new regime where the MDR effects are manifest.

For the $\kappa=1$ case plotted in figure \ref{qresults}  ($n=1,z_c = 10^{21}$) we find that, within the extent of the expected validity of the MDR, the new mass limit $M_{\rm MDR}$ becomes 
\begin{equation}
M_{\rm MDR} = 1.618 \left(\frac{\mu}{2} \right)^2 M_\odot
\end{equation}
about 10 \% larger than the usual, $1.456 M_\odot$, result. 
% which ST give as 1.457, astro text gives as 1.44 and Pad. gives as 1.459
 The case for which $\kappa=-1$ and $\delta z_c = -0.1$ gives 
\begin{equation}
\label{negkapmass}
M_{\rm MDR} = 1.333 \left(\frac{\mu}{2} \right)^2 M_\odot.
\end{equation}
Notice that it is possible to either increase or decrease the Chandrasekhar mass limit.  

The radius of the star is found from $\zeta_R$, $R=r_o \zeta_R$.  As in the usual case, because of the scaling of $z_c$ in $r_o$, the radii vanish as $z_c \rightarrow \infty$.  In the above examples the values for $R\mu$ are $6 \times 10^{-12}$ cm and $8 \times 10^{-12}$ cm, respectively, as anticipated in the qualitative argument.  It is clear from these figures that new physics, such as  the general relativistic correction (the Schwarzchild radius is a few kilometers), dominates well before the MDR effects become important.

\begin{figure}
      \begin{center}
	\includegraphics[scale=1.0]{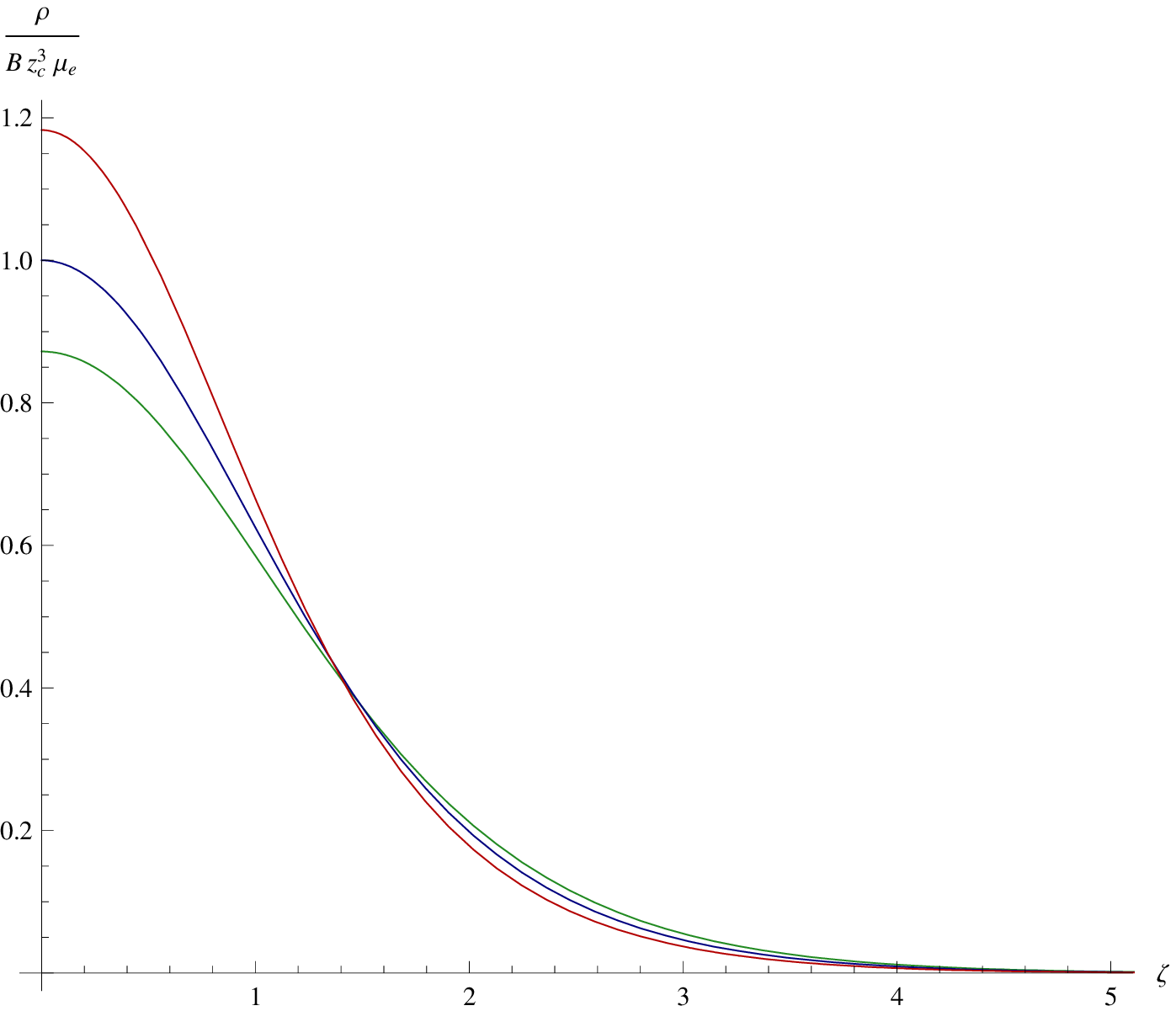}
      \end{center}
\caption{\label{density} The densities $\rho/ (\mu \rho_o z_c^3)$ are shown for $|\delta z_c| = 0.1$.  On the left of the plot the solutions are, from top to bottom $\kappa= -1$ (red), $\kappa= 0$ (blue), and $\kappa= +1$ (green).}
\end{figure}

The density $\rho/(\mu \rho_o) \equiv (x_F(Q)/z_c)^3$ for the solutions of the cases $\kappa = \pm 1$ and $\kappa =0$ ($|\delta z_c| = 0.1$) are plotted in figure \ref{density}.  They are manifestly non-constant.    

The solution space of equilibrium configurations is confined by the regime of validity of the MDR,  $\delta z_c^n  < 1$. Within this limit the $n=1$ case was solved using the exact expression for the physical root $x_F(Q)$ of the cubic equation.   Solutions showed corrected mass limits as the central normalized energy rose above $10^{16}$.  As one might expect, the corrections become significant as $\delta z_c$ increases to 1/2, when the mass limit becomes about $1.94$ solar masses for $\kappa > 0$ and $.84$ for $\kappa <0$.  So as the central energy increased from $\sim 10^{16}$ the MDR effects continued to increase (or decrease) the mass limit to the limits of the effective description at onset of the Planckian regime.

The  $n>1$ solution spaces displayed qualitatively similar effects on the mass limit. Using the approximation of equation (\ref{approx}), the solutions were obtained for $n=2,3,4$. For example, the mass limit for $n=3, \kappa =1, z_c = 3 \times 10^{21}$ becomes $1.557 (\mu/2 )^2 M_\odot$.

In summary, the numerical solutions show that the Chandrasekhar mass limit is altered by about 10/\%due to the modifications in the dispersion relations, increasing for positive $\kappa$ and decreasing for negative $\kappa$.  However, these effects occur at such high central densities that dynamical instabilities occur well before such energies are reached, rendering the MDR effects irrelevant for astrophysics.
 
%\begin{table}
%      \begin{tabular}{l|l|l|l|l|l} \hline
 %     $ \log z_c$ & $\delta z_c^n$ & $\zeta_R$ & $(-\zeta d Q/ d\zeta)_R$ & $ \frac{M}{M_\odot} \left( \frac{2}{\mu} \right)^2$ & $ \frac{R}{\mu} \times 10^7$ (m) \\
 %     \hline 
   %   $n=1$ \\
   %   \hline
    %  2 & $1.4 \times 10^{-20}$ & 7.146 & 2.01715 & 1.455 & $5.55 \times 10^5$ \\
   %   \end{tabular}
%\caption{\label{results} The densities $\rho/ (\mu B z_c^3)$ are shown for the same cases as in figure (\ref{qresults}).  On the left of the plot the solutions are, from top to bottom $\kappa= -1$ (red), $\kappa= 0$ (blue), and $\kappa= +1$ (green).}
%\end{table}

\section{Conclusion}
\label{conclude}

Unlike the threshold calculations, which are rich in new features and lead to strong constraints on the parameters (e.g. \cite{new_crab}), the calculation of the Chandrasekhar mass limit using MDR is straightforward.  The mass limit is raised or lowered according to the sign of the modification. As one can see in the numerical results, the qualitative argument, or from equation (\ref{approx}) for high $z_c$, the effects become important when $\delta z_c^n Q^n \sim 1$ or near the Planck scale, $E/M_P \sim 1$.  This is well-beyond the scale at which other well-known processes occur so these new mass limits are not directly astrophysically relevant. 

This exact analysis corrects the results reported in \cite{camacho}, where it is suggested that current data rule out the $\kappa > 0$ case ($\alpha < 0$ in \cite{camacho}).  In the analysis of \cite{camacho} the MDR correction to the radius, $\delta R$,  is
\begin{equation}
\delta R = \kappa \frac{\hbar}{5 M_P}  \left( \frac{9 \pi M}{m_u} \right)^{1/3} 
\left[1 - \left( \frac{M}{\tilde{M}} \right)^{2/3} \right]^{-3/2}
\end{equation}
where $\tilde{M}^{2/3} = (\hbar/3 \pi G) ( 9 \pi /8 m_u)^{4/3}$.  Hence, as $M \rightarrow \tilde{M}$ the MDR correction grows arbitrarily large.  This observation led Camacho to conclude that current data seems to rule out positive values of $\kappa$.  This conclusion rests from the incorrect assumption that the density is uniform.  As the present analysis shows (see figure \ref{density}) the density is not constant throughout the star.  While the radius does increase with central energies for positive $\kappa$, the actual radii at these energies are so small that other processes such as neutronization and gravitational collapse occur long before the star evolves to that state.

Current observations seem to indicate white dwarfs with smaller radii than expected, at least for iron cores \cite{mathews,provencal}.  The conclusion of \cite{camacho} suggests that solutions with negative $\kappa$ (or $\alpha > 0$), when the white dwarf radius in the MDR case is reduced, lead to a possible explanation for the smaller radii. However as seen in the result reported after equation (\ref{negkapmass}), the model with the realistic density shows significant corrections only at the Planck scale.

The argument rests on fermion statistics and the modified dispersion relation.  Hence, in so far as these relations hold, it would seem that the calculations of the mass limits generalize to frameworks outside the EFT approach.
However, as pointed out in \cite{dsr_chandra}, this neglects the possible changes in the integration measure of the pressure, equation \ref{Pform}.  For instance, in deformed special relativity (DSR) \cite{AC,MS}  the modified dispersion relations are regarded as the invariant of a relativity group with two invariant scales - not only the speed of massless modes, $c$, but also an invariant length or energy\footnote{A particle process threshold analysis of DSR theories does not provide significant constraints on the parameters of the theory.  This is due the fact that in the DSR framework there are no new thresholds and the usual special relativity thresholds are only slightly modified \cite{dsrthresh}.}. As interpreted by Hossenfelder \cite{H} the DSR dispersion relations arise from an effective description of gravitational effects at high energy-density.   The calculation of the neutron star mass in DSR was carried out in \cite{dsr_chandra}. 

So barring modification of statistics or of the measure, the present calculations are valid {\em mutatis mutandis} in a more general context. The new mass limits would appear not to have astrophysical relevance.  Neutronization and the effects of general relativity would still occur at much lower momenta.  Recently there has been work on the impact of MDRs on the structure of astrophysical objets \cite{bz}. 
 
The MDR effects on the Chandrasekhar mass limit may also be interpreted in the context of generalized uncertainty principle.  The non-linear relation between particle momentum in a high energy-density region and the asymptotic region, induces a modification in the Heisenberg uncertainty principle \cite{H}
\begin{equation}
\Delta x \Delta p \geq \frac{\hbar}{2} \left| \ev{ \frac{\partial p}{\partial k} } \right|
\end{equation}
Expanding the modification as  $\partial p/ \partial k = 1+ \kappa (p/M_P)^n$ and following the method of estimating the ground state energy of the hydrogen atom from the uncertainty principle, one may show that the pressure is modified as in equation (\ref{Pform}) (up to a factor of 1/2).  This leads to the same effects as derived in the EFT context.

\begin{acknowledgments}
Thanks to Giovanni Amelino-Camelia, Franz Hinterleitner, and Francis Wilkin for discussions. This work was supported in part by a Research Corporation grant to SAM.
\end{acknowledgments}

\end{document}